\documentclass[prc,showpacs,preprintnumbers,amsmath,amssymb]{revtex4-1}
\usepackage[dvips]{graphicx}
\usepackage{graphics}
\usepackage{bm}

\begin{document}
\title{Effects of ground-state correlations on damping of giant dipole resonances in $LS$ closed shell nuclei}
\author{Mitsuru Tohyama}
\affiliation{Faculty of Medicine, Kyorin University, Mitaka, Tokyo
  181-8611, Japan \email{tohyama@ks.kyorin-u.ac.jp}}
\begin{abstract}
The effects of ground-state correlations on the damping of isovector giant dipole resonances in $LS$ closed shell nuclei $^{16}$O and $^{40}$Ca 
are studied using extended
random-phase-approximation (RPA) approaches derived from the time-dependent density-matrix theory. It is pointed out that unconventional two-body amplitudes of one particle--three hole and three particle--one hole types which
are neglected in most extended RPA theories play an important role in the fragmentation of isovector dipole strength.
\end{abstract}
%\keyword{Extended RPA theory, Isovector dipole giant resonance}
\maketitle
The random phase approximation (RPA) based on the Hartree-Fock (HF) ground state has extensively been used as a standard theory to study giant resonances. RPA describes giant resonances as highly collective states consisting of one particle (p) - one hole (h) excitations.
Most observed giant resonances show strong fragmentation of transition strength, however. For realistic description of giant resonances, therefore, beyond RPA theories that include configurations higher than 1p--1h's are needed. 
The second RPA (SRPA) \cite{srpa} includes the coupling to 2p--2h configurations. The particle-vibration coupling or quasiparticle-phonon models \cite{elena} express 
p--h correlations included in 2p--2h configurations by phonons.
Our extended RPA (ERPA) formulated based on the ground state in the time-dependent density-matrix theory (TDDM) \cite{WC,GT,toh20}
also consists of the coupled equations for one-body and two-body amplitudes as SRPA. Since ERPA is formulated based on a correlated ground state, it
contains the effects of ground-state correlations through the fractional occupation probability $n_\alpha$ of a single-particle state $\alpha$ and the correlated part $C_2$ of a two-body density matrix.
A special feature of ERPA is that the one-body and two-body amplitudes are not restricted to usual 1p--1h and 2p--2h types because the fractional occupation of single-particle states allows us to define
all kinds of one-body and two-body amplitudes such as 1p-1p, 1h-1h, 1p--3h and 3p--1h amplitudes. These unconventional 1p--3h and 3p--1h amplitudes have never been included in the applications of various extended RPA theories 
that incorporate the effects of ground correlations through $n_\alpha$ and $C_2$ \cite{srpa,taka2,robin}. 
In this paper it is demonstrated that the unconventional two-body amplitudes play an important role in the fragmentation of the isovector dipole strength in doubly $LS$ closed shell nuclei $^{16}$O and $^{40}$Ca. 
 
The ground state used in ERPA is given as a stationary solution of the TDDM equations. 
The TDDM equations consist of the coupled equations of motion for the one-body density matrix $n_{\alpha\alpha'}$ 
(the occupation matrix) and the correlated part of the two-body density matrix $C_{\alpha\beta\alpha'\beta'}$
($C_2$). The equations of motion for reduced density matrices form
a chain of coupled equations known as the Bogoliubov-Born-Green-Kirkwood-Yvon (BBGKY) hierarchy and $C_2$ couples to the correlated part $C_3$ of a three-body density matrix.
In this work the BBGKY
hierarchy is truncated by using the following approximation for specific components of $C_3$ \cite{ts14}:
\begin{eqnarray}
C_{\rm p_1p_2h_1p_3p_4h_2}&=&\sum_{\rm h}C_{\rm hh_1p_3p_4}C_{\rm p_1p_2h_2h},
\label{purt1}\\
C_{\rm p_1h_1h_2p_2h_3h_4}&=&\sum_{\rm p}C_{\rm h_1h_2p_2p}C_{\rm p_1ph_3h_4},
\label{purt2}
\end{eqnarray}
where p and h refer to particle and hole states, respectively.
These 2p1h-2p1h and 1p2h-1p2h components of $C_3$ are the leading-order terms in perturbative expansion of $C_3$ using
the Coupled-Cluster-Doubles (CCD)-like ground state wavefunction \cite{ts14}.
The stationary solution of the TDDM equations can be obtained by using either an adiabatic method or a usual gradient method \cite{toh20}.

The ERPA equations for one-body amplitudes $x^\mu_{\alpha\alpha'}$ and two-body amplitudes $X^\mu_{\alpha\beta\alpha'\beta'}$ are derived from
the equation-of-motion approach \cite{s21} assuming the excitation operator 
\begin{eqnarray}
Q^\dag_\mu=\sum_{\alpha\alpha'}x^\mu_{\alpha\alpha'}a^\dag_\alpha a_{\alpha'}+\sum_{\alpha\beta\alpha'\beta'}X^\mu_{\alpha\beta\alpha'\beta'}a^\dag_\alpha a^\dag_\beta a_{\beta'} a_{\alpha'}
\end{eqnarray}
destructs the ground state $|0\rangle$ as $Q_\mu|0\rangle=0$ and excites an excited state $|\mu\rangle$ as $|\mu\rangle=Q^\dag_\mu|0\rangle$. 
Here, $a^\dag_\alpha ~(a_\alpha)$ is the creation (annihilation) operator of a nucleon at a single-particle state $\alpha$.
The equations in ERPA are written in the matrix form 
\begin{eqnarray}
\left(
\begin{array}{cc}
A&B\\
C&D
\end{array}
\right)\left(
\begin{array}{c}
{x}^\mu\\
{X}^\mu
\end{array}
\right)
=\omega_\mu
\left(
\begin{array}{cc}
S_{11}&T_{12}\\
T_{21}&S_{22}
\end{array}
\right)
\left(
\begin{array}{c}
{x}^\mu\\
{X}^\mu
\end{array}
\right),
\label{ERPA1}
\end{eqnarray}
where $\omega_\mu$ is the excitation energy of an excited state $|\mu\rangle$,
$A$, $B$, $C$ and $D$ are the ground-state expectation values of the double commutators between the Hamiltonian and either one-body or two-body excitation operators while 
$S_{11}$, $T_{12}~(=T_{21}^\dag)$ and $S_{22}$ are the ground-state expectation values of the commutators between either one-body or two-body excitation operators.
Each matrix element in Eq. (\ref{ERPA1}) is given explicitly in Ref. \cite{ts08}.
The effects of ground-state correlations are included
in Eq. (\ref{ERPA1}) through $n_{\alpha\alpha'}$ and $C_2$. 
If the HF assumption is made for the ground state,  Eq.(\ref{ERPA1}) is  reduced to
the SRPA equation \cite{srpa}.
The norm matrix consisting of $S_{11}$, $T_{12}$, $T_{21}$ and $S_{22}$ is hermitian by definition, and the matrices $A$, $B$ and $C$ 
satisfy $A=A^\dag$ and $B=C^\dag$ because $n_{\alpha\alpha'}$ and $C_2$ fulfill the stationary conditions \cite{ts08}
\begin{eqnarray}
i\hbar\dot{n}_{\alpha\alpha'}=\langle 0|[a^\dag_{\alpha'}a_\alpha,H]|0 \rangle=0, \\
i\hbar\dot{C}_{\alpha\beta\alpha'\beta'}=\langle 0|[a^\dag_{\alpha'}a^\dag_{\alpha'}a_\beta a_\alpha,H]|0\rangle=0, 
\end{eqnarray}
where $H$ is the total Hamiltonian. However, $D^\dag=D$ is not completely satisfied 
because the approximated $C_3$ (Eqs. (\ref{purt1}) and (\ref{purt2})) does not
fulfill a stationary condition for $C_3$. In the numerical application of ERPA shown below, $D=D^\dag$ is imposed such that $D_{ij}\rightarrow (D_{ij}+D_{ji})/2$, where $i$ and $j$ mean
two-body configurations.

The importance of each two-body configuration in the damping of giant resonances may be estimated from the value of $S_{22}$. When $C_{\alpha\beta\alpha'\beta'}$ is neglected for simplicity, 
the diagonal element of $S_{22}$ is given by \cite{ts08,toh18}
\begin{eqnarray}
S_{22}(\alpha\beta\alpha'\beta':\alpha\beta\alpha'\beta')&=&(1-n_\alpha)(1-n_\beta)n_{\alpha'}n_{\beta'}
%\nonumber \\
-n_\alpha n_\beta (1-n_{\alpha'})(1-n_{\beta'}),
\end{eqnarray}
where $n_{\alpha\alpha'}=\delta_{\alpha\alpha'}n_\alpha$ is assumed.
In the case of the HF ground state where $n_\alpha=0$ or 1, $S_{22}$ is not vanishing only for the 2p--2h and 2h--2p configurations:
$S_{22}$ is 1 $(-1)$ for the 2p--2h (2h--2p) configurations $X^\mu_{\rm pp'hh'}$ ($X^\mu_{\rm hh'pp'}$). When the single-particle states are fractionally occupied, all two-body configurations
can have non-vanishing values of $S_{22}$. Let us assume that $n_{\alpha}=\Delta$ for a particle state and $n_{\alpha}=1-\Delta$
for a hole state
independently of $\alpha$ and that $\Delta$ is small. Then, $S_{22}\approx 1-4\Delta$ for the 2p--2h configurations and 
the 3p--1h (1h--3p) and 1p--3h (3h--1p) configurations, $X^\mu_{\rm pp'p''h}$ ($X^\mu_{\rm hpp'p''}$) and $X^\mu_{\rm phh'h''}$ ($X^\mu_{\rm hh'h''p}$), have $S_{22}\approx \Delta (-\Delta)$. 
The values of $S_{22}$ for other configurations are
of higher order of $\Delta$. This suggests that the 3p--1h (1h--3p) and 1p--3h (3h--1p) states are the next order configurations to be included
when the effects of ground-state correlations are considered.
That these configurations play a particularly important role when ground states are correlated is also borne out by the fact that their inclusion to
the RPA operator leads to a destructor which annihilates exactly the CCD ground state \cite{st16}.

With the use of a small single-particle space needed to calculate $n_{\alpha}$ and $C_{2}$ and to define the 3p--1h and 1p--3h configurations, 
the fragmentation of the isovector dipole strength in $^{16}$O is studied and the result in ERPA is compared
with that in exact diagonalization approach (EDA).
The occupation probability $n_{\alpha}$ and $C_{2}$ in $^{16}$O are
calculated within TDDM using the $1p_{1/2}$, $1p_{3/2}$ and $1d_{5/2}$ states for both protons and neutrons. 
The single-particle energies and wavefunctions are calculated using the Skyrme III force \cite{skIII}.
A simplified interaction that contains only the $t_0$ and $t_3$ terms of the Skyrme III force is used as the residual interaction to facilitate numerical calculations \cite{toh07}.
The ground state is obtained by using the adiabatic method, which is explained in Ref. \cite{toh21} in some detail.
The 3p--1h and 1p--3h (and 1h--3p and 3h--1p) amplitudes in ERPA are defined by using the same single-particle states as those used in the ground-state calculation. 
Note that in the small single-particle space considered here, $X^\mu_{\alpha\beta\alpha'\beta'}$ for electric dipole states cannot have the 2p--2h and 2h--2p components. 
In the so-called $M$ scheme used in this work for $^{16}$O the numbers of the matrix elements 
of $C_2$, $C_3$ and $X^\mu_{\alpha\beta\alpha'\beta'}$ are about 8500, $1.7\times 10^6$ and 8300, respectively, whereas the number of configurations in EDA which also uses the $M$ scheme is about $4.6\times 10^4$.

The occupation probabilities calculated in TDDM for $^{16}$O are shown in Table \ref{tab1}. 
The results in EDA which are obtained by using the same single-particle states and interaction as those used in TDDM
are given in parentheses. The results in TDDM agree well with the EDA results.
The deviation of $n_\alpha$ from the HF values ($n_\alpha=1$ or 0) is close to 10 \%, 
indicating that the ground state of $^{16}$O is highly correlated
as other calculations \cite{agassi,adachi,utsuno} have already suggested. The values in brackets
will be explained below.
\begin{table}
\caption{Single-particle energies $\epsilon_\alpha$ and occupation probabilities 
$n_{\alpha}$ calculated in TDDM for $^{16}$O. The results in EDA are given in parentheses.
The values in brackets give the results of the TDDM calculation that neglects $C_3$ and includes only the 2p--2h and 2h--2p
components of $C_2$.}
\begin{center}
\begin{tabular}{c rr rr} \hline
 &\multicolumn{2}{c}{$\epsilon_\alpha$ [MeV]}&\multicolumn{2}{c}{$n_{\alpha}$}\\ \hline 
orbit & proton & neutron  & proton & neutron  \\ \hline
$1p_{3/2}$ & -18.2 & -21.8 & 0.913 (0.910) [0.907]& 0.913 (0.910) [0.908]  \\
$1p_{1/2}$ & -12.0 & -15.6 & 0.889 (0.883) [0.879] & 0.887 (0.883) [0.879]  \\
$1d_{5/2}$ & -3.8 & -7.2 & 0.095 (0.099) [0.102] & 0.096 (0.099) [0.102]  \\\hline
\end{tabular}
\label{tab1}
\end{center}
\end{table}

\begin{figure} 
\begin{center} 
\includegraphics[height=6cm]{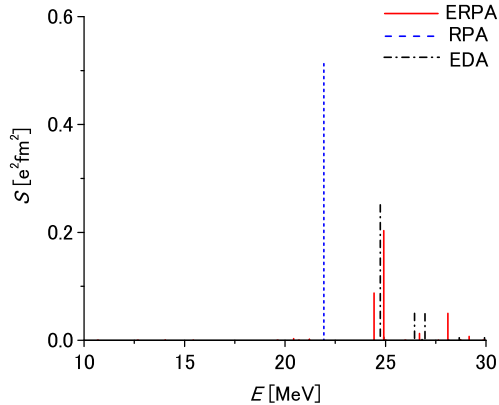}
\end{center}
\caption{Isovector dipole strength distributions calculated in ERPA (solid lines), RPA (dashed line) and EDA (dot-dashed lines) for $^{16}$O.} 
\label{oe11} 
\end{figure}

In Fig. \ref{oe11} the isovector dipole strength distributions calculated in ERPA (solid lines),  RPA (dashed line) and EDA (dot-dashed lines) are shown for $^{16}$O. 
The peak at 21.9 MeV in RPA consists of the $1p_{3/2}\rightarrow 1d_{5/2}$ transitions and corresponds to the isovector giant dipole resonance (GDR).
The dipole strength in ERPA is split into several states around $E=25$ MeV due to the coupling to the 3p--1h and 1p--3h configurations: 
The contributions of the 3h--1p and 1h--3p configurations are small because they have negative energies and are energetically separated from GDR.
As mentioned above, there are no dipole states consisting of the 2p--2h and 2h--2p configurations 
in the small single-particle space used here. Since the unperturbed energies of the 3p--1h and 1p--3h configurations are smaller than the GDR energy, 
the isovector dipole strength is shifted upward in ERPA due to the coupling 
to these two-body configurations. The EDA strength is also split into several states, which agrees with the ERPA result though there is some difference in the location and strength of each state.
The results in Fig. \ref{oe11} demonstrate the importance of the unconventional two-body amplitudes in the study of fragmentation of GDR.

The inclusion of $C_3$ in the equations in TDDM and ERPA is numerically quite challenging except for simple cases as shown above or solvable models \cite{st16}.
In the previous applications to giant resonances in oxygen and calcium isotopes \cite{toh07,toh18} simpler approximations have been used: 
The ground states are calculated by using a simple truncation scheme of the BBGKY hierarchy that neglects $C_3$ and keeps only the 2p--2h and 2h--2p components of $C_2$,
and the excited states are obtained from
the small amplitude limit of TDDM (STDDM). In the following the validity of this simple approach for $^{16}$O is investigated and also
the STDDM result for $^{40}$Ca is presented. First let us explain a relation of ERPA and STDDM.
From the small amplitude limit of the TDDM equations that do not include $C_3$, the coupled equations in STDDM are obtained 
for the one-body transition amplitudes $\tilde{x}^\mu_{\alpha\alpha'}=\langle 0|a^\dag_{\alpha'}a_\alpha|\mu\rangle$ and the two-body transition amplitudes 
$\tilde{X}^\mu_{\alpha\beta\alpha'\beta'}=\langle 0|a^\dag_{\alpha'}a^\dag_{\beta'}a_\beta a_\alpha|\mu\rangle$. They are written in matrix form as
\begin{eqnarray}
\left(
\begin{array}{cc}
a&b\\
c&d
\end{array}
\right)\left(
\begin{array}{c}
\tilde{x}^\mu\\
\tilde{X}^\mu
\end{array}
\right)
=\omega_\mu
\left(
\begin{array}{c}
\tilde{x}^\mu\\
\tilde{X}^\mu
\end{array}
\right).
\label{STDDM}
\end{eqnarray}
The matrices $a$, $b$, $c$ and $d$ are explicitly given in Ref. \cite{ts08}.
With the use of 
\begin{eqnarray} 
\left(
\begin{array}{c}
\tilde{x}^\mu\\
\tilde{X}^\mu
\end{array}
\right)=
\left(
\begin{array}{cc}
S_{11}&T_{12}\\
T_{21}&S_{22}
\end{array}
\right)
\left(
\begin{array}{c}
{x}^\mu\\
{X}^\mu
\end{array}
\right),
\end{eqnarray}
Eq. (\ref{STDDM}) can be transformed to another matrix form similar to Eq. (\ref{ERPA1}) as 
\begin{eqnarray}
\left(
\begin{array}{cc}
A&B\\
C'&D'
\end{array}
\right)\left(
\begin{array}{c}
{x}^\mu\\
{X}^\mu
\end{array}
\right)
=\omega_\mu
\left(
\begin{array}{cc}
S_{11}&T_{12}\\
T_{21}&S_{22}
\end{array}
\right)
\left(
\begin{array}{c}
{x}^\mu\\
{X}^\mu
\end{array}
\right),
\label{STDDM1}
\end{eqnarray}
where $A=aS_{11}+bT_{21}$, $B=aT_{12}+bS_{22}$, $C'=cS_{11}+dT_{21}$ and $D'=cT_{12}+dS_{22}$.
The matrices $A$ and $B$ are the same as those in Eq. (\ref{ERPA1}) but $D'\neq D$.
In order to express $D$ that is the ground-state expectation value of the double commutators between the Hamiltonian and two-body excitation operators, we need
an additional term $eT_{32}$ where $e$ depicts the coupling of the two-body transition amplitudes to the three-body transition amplitudes and $T_{32}$ is the expectation value of the commutator between
the three-body and two-body excitation operators \cite{ts08}. When $C_3$ is included, $T_{13}$ that is the expectation value of the commutator between the one-body and three-body
excitation operators is not vanishing and $C$ in Eq. (\ref{ERPA1}) is obtained by adding a term $eT_{31}$ to $C'$. Thus ERPA includes the three-body effects that are not considered in STDDM.
The small amplitude limit of the approximate $C_3$ given by Eqs. (\ref{purt1}) and (\ref{purt2}) gives similar additional terms but they cannot fully express $eT_{32}$ and $eT_{31}$ because the matrix elements of $C_3$ are
restricted to the 2p1h-2p1h and 2h1p-2h1p types.

\begin{figure} 
\begin{center} 
\includegraphics[height=6cm]{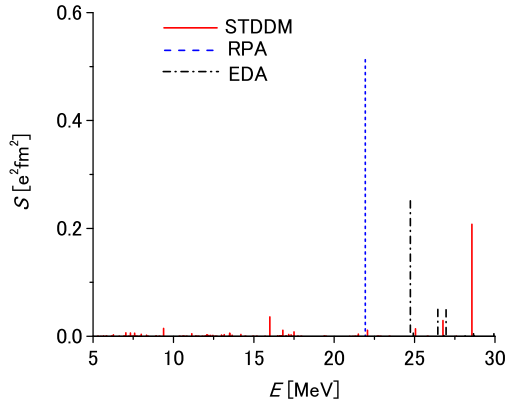}
\end{center}
\caption{Isovector dipole strength distributions calculated in STDDM (solid lines), RPA (dashed line) and EDA (dot-dashed lines) for $^{16}$O.} 
\label{oe1} 
\end{figure}

The occupation probabilities $n_\alpha$ in $^{16}$O calculated using the simple TDDM approach that neglects $C_3$ and keeps only the 2p--2h and 2h--2p components of $C_2$ are given in Table \ref{tab1} (the values in brackets). 
The number of the 2p--2h and 2h--2p components of $C_2$ is about 2000 in the $M$ scheme.
The occupation probabilities in the simple TDDM also show good agreement with the EDA results.
In Fig. \ref{oe1} the isovector dipole strength distribution calculated in STDDM (solid lines) is compared with the results in RPA (dashed line) and EDA (dot-dashed lines) for $^{16}$O. 
The strength distribution in STDDM above the GDR energy (21.9 MeV) agrees with the EDA and ERPA results in the sense that the dipole strength is split into several states
though the main peak is slightly shifted upward as compared with the ERPA and EDA results. 
STDDM shows some small strength distribution below GDR, which differs from 
the EDA and ERPA results: EDA and ERPA also have several states below GDR but their transition strengths are invisible in the scale of Figs. \ref{oe11} and \ref{oe1}. 
In ERPA the $eT_{32}$ term which includes the effects of self-energy contributions to the 3p--1h and 1p--3h configurations plays an important role in 
increasing the energies of these configurations and suppressing the strength distribution in the low energy region,
while the $eT_{31}$ term which modifies the coupling of $x^\mu$ to $X^\mu$ has little effects of suppressing the number of states below GDR.
From the above calculation it is found that STDDM can be used to study the fragmentation of GDR though it cannot properly treat the strength distribution in low energy region. 

Next the results of STDDM calculations for $^{40}$Ca are presented. Similarly to the case of $^{16}$O the ground state is calculated by using the small single-particle space consisting of 
the $2s_{1/2}$, $1d_{3/2}$, $1d_{5/2}$ and $1f_{7/2}$ states for both protons and neutrons. The simple interaction consisting of the $t_0$ and $t_3$ terms of the Skyrme III is used as the residual interaction.
In the $J$ scheme used for $^{40}$Ca \cite{toh18} the number of the matrix elements of $C_2$ is 336.
In this ground-state calculation the gradient method is used to solve the TDDM equations \cite{toh18}.
The results of the ground-state calculation for $^{40}$Ca that neglects $C_3$ and keeps only the 2p--2h and 2h--2p components $C_2$ are given in Table \ref{tab2}.
The occupation probabilities obtained are comparable to the results of other calculations \cite{agassi,adachi}.
\begin{table}
\caption{Single-particle energies $\epsilon_\alpha$ and occupation probabilities 
$n_{\alpha\alpha}$ calculated in TDDM for $^{40}$Ca. Here, $C_3$ is neglected and only the 2p--2h and 2h--2p components of $C_2$ are included.}
\begin{center}
\begin{tabular}{c rr rr} \hline
 &\multicolumn{2}{c}{$\epsilon_\alpha$ [MeV]}&\multicolumn{2}{c}{$n_{\alpha}$}\\ \hline 
orbit & proton & neutron  & proton & neutron  \\ \hline
$1d_{5/2}$ & -15.6 & -22.9 & 0.923 & 0.924  \\
$1d_{3/2}$ & -9.4 & -16.5 & 0.884 & 0.884  \\
$2s_{1/2}$ & -8.5 & -15.9 & 0.846 & 0.846   \\ 
$1f_{7/2}$ & -3.4 & -10.4 & 0.154 & 0.154  \\\hline
\end{tabular}
\label{tab2}
\end{center}
\end{table}

\begin{figure} 
\begin{center} 
\includegraphics[height=6cm]{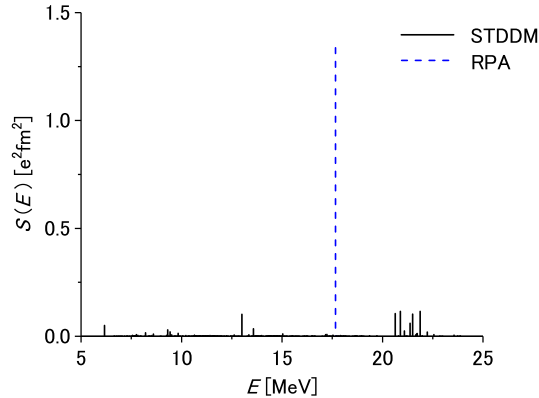}
\end{center}
\caption{Isovector dipole strength distributions calculated in STDDM (solid lines) and RPA (dashed line) for $^{40}$Ca.}
\label{cae1} 
\end{figure}

The isovector dipole strength distribution in STDDM for $^{40}$Ca is first calculated using the same small single-particle space as that used for the 
ground-state calculation.  In this truncated single-particle space there are no dipole states consisting of the 2p--2h and 2h--2p configurations.
The numbers of the matrix elements of $X^\mu_{\alpha\beta\alpha'\beta'}$ is 2048 in the $J$ scheme.
The STDDM result (solid lines) is shown in Fig. \ref{cae1}. 
The result in RPA is depicted with 
the dashed line in Fig. \ref{cae1}.
Even in the small-single particle space used for $^{40}$Ca it is hard to perform an EDA calculation which uses $M$-scheme, 
and the EDA results are not given in Fig. \ref{cae1}.
The peak at 17.7 MeV in RPA consists of the $1d_{5/2}\rightarrow 1f_{7/2}$ transitions and corresponds
to GDR.
As in the case of $^{16}$O the dipole strengths are fragmented in STDDM due to the coupling to the 3p--1h and 1p--3h configurations. 
Since the unperturbed 3p--1h and 1p--3h states
are located around GDR, the STDDM dipole strength above GDR is more fragmented in $^{40}$Ca than in $^{16}$O. 

\begin{figure} 
\begin{center} 
\includegraphics[height=6cm]{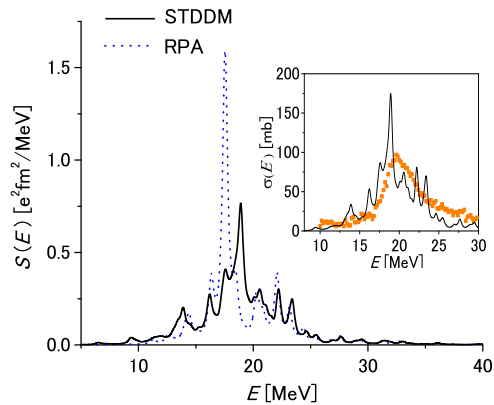}
\end{center}
\caption{Isovector dipole strength functions calculated in STDDM (solid line) and RPA (dotted line) for $^{40}$Ca. 
The distributions are smoothed with an artificial width $\Gamma=0.5$ MeV.
In the inset the photoabsorption cross section calculated from the strength function in STDDM (solid line) is compared with experimental data (squares) \cite{arhens}.}
\label{cae11} 
\end{figure}

The result of a more realistic STDDM calculation that includes a large number of single-particle states for $x^\mu_{\alpha\alpha'}$ and thus can be compared with experiment
is presented below.
The one-body amplitudes ${x}^\mu_{\alpha\alpha'}$ are defined with a large number of single-particle states including those in the 
continuum: The continuum states are discretized by confining the wavefunctions in a sphere with radius 15 fm and all 
the single-particle states with $\epsilon_\alpha\le 50$ MeV and 
$j_\alpha\le 9/2 \hbar$ are included. 
Since the residual interaction is not consistent with the effective interaction used in the calculation of the single-particle states,
it is necessary to reduce the strength of the residual interaction in the one-body channels when the large single-particle space is used for ${x}^\mu_{\alpha\alpha'}$. 
The reduction factor $f$ is determined so that the spurious mode corresponding
to the center-of-mass motion comes at zero excitation energy in RPA. It is found that $f=0.66$ \cite{toh18}. This factor is used in the $a$, $b$ and $c$ parts of Eq. (\ref{STDDM}).
The two-body amplitudes are defined by using the same single-particle states as those used in the ground-state calculation. It is stressed again that 
in this small single-particle space the two-body amplitudes for isovector dipole states only have the 1p--3h (3h--1p) and 3p--1h (1h--3p) components.
The result of the STDDM calculation for the isovector dipole excitation in $^{40}$Ca is shown in Fig. \ref{cae11}
with the solid line. The dotted line depicts the result in RPA. The distributions are smoothed with an artificial width $\Gamma=0.5$ MeV.
Figure \ref{cae11} shows that the inclusion of the 3p--1h and 1p--3h configurations in STDDM significantly increases the fragmentation of the dipole strength. 
The fragmentation of GDR in STDDM is comparable to the results in other theoretical approaches such as the particle-phonon coupling models \cite{tsel,egoro} that include much larger two-body configurations than
the STDDM calculation. The photoabsorption cross section calculated in STDDM can also be compared with
the experimental data (squares) \cite{arhens} as shown in the inset of Fig. \ref{cae11}. Better agreement with the experiment would be obtained by the inclusion of 
2p--2h configurations as is done in 
large scale SRPA calculations \cite{gamb1,gamb2} and also by the improvement in the treatment of continuum states.

In summary, 
the effects of ground-state correlations on the damping of the isovector giant dipole resonances in $LS$ closed shell nuclei $^{16}$O and $^{40}$Ca 
were studied using extended RPA approaches derived from the time-dependent density-matrix theory. 
It was pointed out that the unconventional two-body amplitudes of one particle--three hole and three particle--one hole types which
are neglected in most extended RPA theories play an important role in the fragmentation of the dipole strength because the ground states of these nuclei are highly correlated.
Our results suggest that in extended RPA studies of GDR's in these nuclei such unconventional configurations 
should also be included in addition to conventional two--particle two--hole configurations. 

\let\doi\relax

\end{document}